\documentclass[final,reqno,conference]{IEEEtran} 
\usepackage{amsmath,amssymb,amsthm} \usepackage{graphicx,color} 
\usepackage{bbm}
\usepackage{mathrsfs}

\usepackage{subcaption}
\usepackage{epstopdf}

\DeclareMathOperator{\argmin}{argmin}

\title{Optimal Policies of Advanced Sleep Modes for Energy-Efficient 5G networks}
\author{\IEEEauthorblockN{Fatma Ezzahra Salem$^{a,b}$, Tijani Chahed$^{b}$, Eitan Altman$^{c}$, Azeddine Gati$^{a}$, Zwi Altman$^{a}$ }
\IEEEauthorblockA{$^{a}$Orange Labs, Ch\^atillon, France 
\\$^{b}$Institut Polytechnique de Paris, Telecom SudParis, UMR CNRS SAMOVAR, Evry, France
\\$^{c}$INRIA, Sophia-Antipolis, France
\\Email: \{fatma.salem,zwi.altman,azeddine.gati\}@orange.com  \\ tijani.chahed@telecom-sudparis.eu \hspace{0.3cm} eitan.altman@inria.fr}\vspace{0.01cm}}  

\IEEEoverridecommandlockouts
\IEEEpubid{\makebox[\columnwidth]{978-1-7281-2522-0/19/\$31.00~\copyright{}2019 IEEE \hfill}\hspace{\columnsep}\makebox[\columnwidth]{ }}

\begin{document}

\maketitle

\begin{abstract} 
We study in this paper optimal control strategy for Advanced Sleep Modes (ASM) in 5G networks. ASM correspond to different levels of sleep modes ranging from deactivation of some components of the base station for several micro-seconds to switching off of almost all of them for one second or more. ASMs are made possible in 5G networks thanks to the definition of so-called lean carrier radio access which allows for configurable signaling periodicities. We model such a system using Markov Decision Processes (MDP) and find optimal sleep policy in terms of a trade-off between saved power consumption versus additional incurred delay for user traffic which has to wait for the network components to be woken-up and serve it. Eventually, for the system not to oscillate between sleep levels, we add a switching component in the cost function and show its impact on the energy reduction versus delay trade-off. \\

\end{abstract} 

\section{Introduction}
Traditionnally, mobile networks were designed in such a way so as to provide higher data rates, better coverage and ubiquitous connectivity. They have to be always available and well-dimensioned in order to ensure the best Quality of Service (QoS) even in peak hours and mass-event scenarios. This may lead to an over-dimensioned and under-utilized network particularly when the traffic demand is low, such is the case during night hours. 

Energy consumption of the network is composed of two components: i. a fixed one, which is mainly due to the system architecture and includes power consumption of control signals and backhaul infrastructure as well as load-independent consumption of baseband processors \cite{Ref00} and ii. a variable, load-dependent one, which depends on the transported traffic. Over-provisioning of the network as well as low load periods translate into significant, and unnecessary, energy consumption, due to the fixed component. Sleep modes, which consist in shutting down the base station for a certain period of time, are an efficient way to handle this component \cite{Ref011}  and are the main focus of the present work. 



During the last decade, the concern about energy consumption of mobile networks triggered a particular attention from industry as well as academia and research institutes. Two major motivations reside behind this increasing interest of building \textit{Green} mobile networks: on the one hand, reducing the operator's operational expenditures (OPEX), and on the second hand, preserving the environment by reducing the CO$_2$ footprint. A new Key Performance Indicator (KPI) was defined: Energy-Efficiency (EE), measured in bit/Joule, that expresses the amount of information (in bits) transmitted per joule of consumed energy \cite{Ref01} and that represents one of the main requirements in the design of 5G networks \cite{Ref02}. Several techniques can enhance the EE of the network in certain cases such as virtualization \cite{Ref03}, device to device communications \cite{Ref04}, local caching \cite{Ref05}, etc. Many projects were launched in order to define strategies and mechanisms that would help to achieve the target of building more energy-efficient networks, such as EARTH \cite{Ref06}, 5GrEEn \cite{Ref07}, GreenTouch \cite{Ref08}, SooGreen \cite{Ref09}, etc.

We focus in this work on a sleep mode feature dedicated to 5G networks called \textit{Advanced Sleep Modes} (ASM). It consists in a progressive deactivation of the base station's components according to the time needed by each of them to shut down then reactivate again. According to this transition time, four levels of sleep modes have been defined \cite{Ref012}. Deeper sleep levels allow more energy saving but induce larger delays for the users who arrive to the network and who need to wait longer for the components to be reactivated. Hence, a trade-off between these two metrics: energy saving versus delay, has to be found. 

In \cite{Ref012_00} \cite{Ref012_01}, the authors propose several deactivation strategies using the different ASM levels. They assess their performance depending on the periodicity of the control signals sent by the base station. Their approches consist on putting the base station into sleep gradually starting from the lightest level (only few components are deactivated) to the deepest possible level. A fixed trajectory of deactivation is then imposed in each idle period (when there is no data transmission). Authors in \cite{Ref012_02} allow switching between the different sleep levels if no user request occurs during the idle period. In their work, performance is discussed in terms of energy savings, while the impact on the latency was not investigated. We presented in a previous work \cite{Ref012_1} a management strategy to orchestrate the different ASM levels based on reinforcement learning addressing the trade-off between energy reduction and delay. We however assumed a pre-defined order for the sleep levels: from deepest to lightest. 

Our aim in this work is to study the more general case where we do not impose a pre-defined path for the sleep levels. At each decision point, the base station is free to decide the next sleep level to go to. We do so based on Markov Decision Processes (MDPs) and derive the optimal sleep policy as well as the resultant power saving versus delay performance. The latter may correspond to oscillations  between the different sleep levels. In order to prevent these oscillations, we introduce an additional switching cost and we study its impact on the optimal policy and performance. We use in this work a similar modeling approach as in \cite{Ref1} and extend it to the case of ASMs.      


The remainder of this paper is structured as follows: Section \ref{Section2} presents the 5G New Radio (NR) compared to the case of 4G networks and describes the ASMs' characteristics. Section \ref{Section3} introduces our system model. Section \ref{Section4} presents the MDP-based analysis. Section \ref{Section5} presents some numerical applications. Finally, Section \ref{Section6} concludes the paper and gives some perspectives for future works.

\section{5G and ASM}
\label{Section2}
\subsection{4G legacy}
In 4G networks, control signals are sent frequently by the base station. For instance Cell-specific Reference Signals (CRS) are broadcasted during 4 OFDM symbols out of 14 (depending on MIMO configuration), Primary and Secondary Synchronization Signals (PSS and SSS, respectively) are transmitted in the first and fifth sub-frames of the LTE radio frame, and the Physical Broadcast CHannel (PBCH) is sent in the first sub-frame. Such frequency of the signaling bursts does not allow to put the base station into sleep for a long period. It allows only to shut down the Power Amplifier (PA) and part of RF subsystem during idle symbols. This technique is called Micro-Discoutinous Transmission ($\mu$DTX) \cite{ref013}. 

\subsection{Energy saving opportunities with 5G}
\subsubsection{Lean carrier design}


It has been agreed in 3GPP \cite{Ref014} that the broadcast signals sent by the base station will be grouped in blocks, named Synchronisation Signal Blocks (SSB), and sent with a periodicity that can be adjusted by the network operator. It can take values in the range 5, 10, 20, 40, 80 and 160 ms. With these values of signaling periodicities, we can use deeper sleep levels than the $\mu$DTX used in 4G.
\subsubsection{5G Stand-Alone and non Stand-Alone}
Two different configurations are possible in 5G NR: 5G Stand-Alone (SA) and 5G Non Stand-Alone (NSA) \cite{3GPP}.

In 5G NSA, the 5G deployment depends on existing LTE network for control functions, while 5G NR is used only for user plane (data transmissions). For 5G SA, the 5G cells are used for both signalling and data transmissions. This means that SA configuration does not allow a sleep duration larger than 160 ms as this is the maximum periodicity allowed for the control signals according to the lean carrier design.

\subsection{Advanced Sleep Modes}
ASM correspond to a gradual deactivation of the different components of the base station depending of their transition times, i.e., the durations of deactivation and activation summed up together. Based on this criteria, four different levels have been defined in \cite{Ref012}: \vspace{0.2cm} \\
\textbf{- SM$_1$:} corresponds to the shortest Sleep Mode (SM) with a duration equal to one OFDM symbol, i.e., 71$\mu s$. Some of the base station's components are disabled, for instance the PA. \vspace{0.2cm} \\
\textbf{- SM$_2$:} this medium level is in the scale of one sub-frame or Transmission Time Interval (TTI) of 1$ms$. More components are deactivated than in $SM_1$. \vspace{0.2cm} \\
\textbf{- SM$_3$:} the third level has a duration of a frame of 10$ms$. In this mode, most of the components are deactivated: all the components of the digital baseband and analog front-end (both Rx and Tx) except the clock generator. \vspace{0.2cm} \\
\textbf{- SM$_4$:} it corresponds to the standby mode with a minimal duration of 1$s$. The base station is out of operation during this mode but the backhaul remains active so as to be able to re-activate it.

The different characteristics of the ASMs are summarized in Table \ref{Table1}. 

For a single sleep mode, we consider that the OFF period is the sum of the deactivation period, the minimum sleep period and the reactivation period.

    \begin{table}[ht]
    \small
    \renewcommand{\arraystretch}{1.3}
    \caption{Advanced Sleep Modes characteristics}
    \label{Table1}
    \centering
    \begin{tabular}{|c|c|c|c|}
    \hline
    \textbf{Sleep} & \textbf{Deactivation} & \textbf{Minimum} & \textbf{Activation} \\
     \textbf{level} & \textbf{duration} & \textbf{sleep duration} & \textbf{duration} \\
         \hline
    \textbf{SM$_1$} & 35.5 $\mu$s & 71 $\mu$s & 35.5 $\mu$s \\
	\hline
	    \textbf{SM$_2$} & 0.5 ms & 1 ms & 0.5 ms \\
	\hline
	    \textbf{SM$_3$} & 5 ms & 10 ms & 5 ms \\
	\hline
	    \textbf{SM$_4$} & 0.5 s & 1 s & 0.5 s \\
	\hline 
    \end{tabular}
    \end{table}
  
In 5G NSA architecture, all the ASM levels are allowed while in 5G SA, we can allow only the first three levels since the duration of SM$_4$ is larger than the SSB periodicity. We focus in this work on the first case.
\section{System Model}
\label{Section3}

\subsection{System description} 
We focus in this work on the downlink. We consider a sleep mode approach where we shut down both the transmitters and the receivers in the base station. The base station cannot listen to incoming traffic during sleep periods. If a request occurs during this period, it is put in a buffer. The base station wakes up from time to time in order to check the status of the buffer, i.e., if it is empty or containing some packets waiting. In the former case, the base station decides to continue its sleep, while in the latter, it has to stay awake and serve the waiting users. Once the service is complete, the base station can go again into sleep. \\

As in \cite{Ref1}, we consider a generic idle period $I$ in which the base station goes into repeated sleep periods that may have different durations. After each sleep period, the base station wakes up to check if there are buffered packets. Unlike \cite{Ref1} we consider in this work several sleep depths, each with a given duration corresponding to the above mentioned ASM levels and an associated power consumption. 


Let $\tau$ be the inactivity period, i.e., the time between the start of the first sleep period and the arrival of the first user during the idle period. $\tau$ is a random variable with a probability density function denoted by $f_{\tau}(t), t \geq 0.$ We assume that $\tau$ is hyper-exponentially distributed as idle periods have been shown to be heavy tailed \cite{Ref1} \cite{Ref2} \cite{Ref3}. 

Thus, 
	\begin{equation} 
		f_{\tau}(t) = \sum_{i=1}^n q_i \lambda_i exp(-\lambda_it),   \text{  }\sum_{i=1}^n q_i =1
	\end{equation}
where $n$ are the phases of the hyper-exponential distribution and $\lambda = (\lambda_1, ..., \lambda_n)$ and $q = (q_1,...q_n)$ its parameters.

\begin{figure} [ht]
\centering
\includegraphics[width=9.2cm]{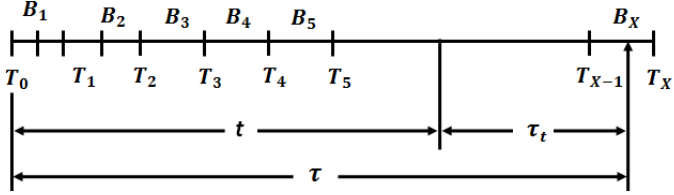}
\caption {Generic inactivity period} 
\label {Fig1}
 \end{figure}

Let $T_k$ denote the time at the end of the $k^{th}$ sleep period and $B_k$ its duration, $k \in \mathbbm{N}^*$. The time at the start of the first sleep period is denoted $T_0$. Hence, $T_k = \sum_{i=1}^{k} B_i$ and $T_X$ corresponds to the end of the generic idle period. This scheme is illustrated by Figure \ref{Fig1}. 
 
At the end of each sleep period $T_k$, the base station checks if there was a user request during period $B_k$. If the buffer is empty, the base station chooses the next sleep level for the next period. Note that it can stay in the current sleep level or switch to another one (deeper or lighter). 

We consider that each sleep period $k$ begins by a deactivation period and is followed by a sleep period and then a warm up period in order to activate the base station. For the sake of simplification, we assume that the power consumed during all these three sleep mode mechanisms is the same and equal to $P_l$, where $ l$ is the sleep mode level during this period $k$. If the base station finds that the buffer contains at least one request, the idle period is finished and the base station starts serving the waiting users. A delay is incurred since the users were buffered until the base station wakes up for checking. This buffering duration depends on the length of the $k^{th}$ sleep period and its type. The first service request will be delayed by $T_X - \tau$.
 \subsection{Cost function}

We define a cost function $V$ as a weighted sum of the total energy consumption and the added delay resulting from the sleep policy as follows:
	\begin{equation}
		V := \mathbb{E}[\epsilon_1(T_X-\tau) + \epsilon_2 T_X\sum_{l=1}^{L}w_l P_l]
	\end{equation}
where $\epsilon_1$ is a normalized weight ($\in [0,1]$) that denotes the importance given to the delay and $\epsilon_2 = 1 - \epsilon_1$ represents the weight given to the energy consumption, $w_l$ is the proportion of time during which sleep mode $l$ was chosen and $P_l$ is the corresponding power consumption. Our target is to find the policy which minimizes the cost function $V$. 

$\epsilon_1$ and $\epsilon_2$ should be taken in such a way so as to satisfy the network operator's objective regarding energy reduction and QoS requirement in terms of delay constraints which depends on the different use cases in 5G netwroks and whether they are delay-sensitive or not. The network operator can impose thresholds on both metrics: energy reduction and delay, and the values of $\epsilon_1$ and $\epsilon_2$ have to be tuned accordingly.

$V$ can also be written as follows: 
		\begin{equation}
		V = -\epsilon_1 \mathbb{E}[\tau] + \eta \mathbb{E}[T_X] 
	\end{equation}
where $\eta = \epsilon_1 + \epsilon_2 \sum_{l=1}^{L}w_l P_l$.  \\ 

For a hyper-exponential distributed off-time $\tau$, we have: 
\begin{equation}
\mathbb{E}[\tau] = \sum_{i=1}^{n} \frac{q_i}{\lambda_i}
\end{equation}

\begin{equation}
\mathbb{E}[T_X] = \sum_{k=0}^{\infty}\sum_{i=1}^{n} q_i  \mathcal{T}_k(\lambda_i)\mathbb{E}[B_{k+1}]
\end{equation}
where $\mathcal{T}_k(\lambda_i) = \mathbb{E}[exp(-\lambda_i T_k)]$. \\

Then, $V$ can be expressed as follows: 
\begin{equation}
V = -\epsilon_1 \mathbb{E}[\tau] + \eta \sum_{k=0}^{\infty}\sum_{i=1}^{n} q_i  \mathcal{T}_k(\lambda_i)\mathbb{E}[B_{k+1}]
\end{equation}

\subsection{Distribution of the conditional residual off-time:}
Let $\tau_t$ denote the conditional residual off-time at time $t$. The tail of $\tau_t$ can be written as follows: 
\begin{equation}
P(\tau_t > a) = P(\tau > t+a)|\tau>t) = \frac{P(\tau > t+a)}{P(\tau > t)} 
\end{equation}

\begin{equation}
= \frac{\sum_{i=1}^{n} q_i exp(- \lambda_i t) exp(- \lambda_i a) }{\sum_{j=1}^{n} q_j exp(- \lambda_j t) }
\end{equation}

\begin{equation}
= \sum_{i=1}^{n} g_i(\textbf{q},t)exp(- \lambda_i a)
\end{equation}

where: 
\begin{equation}
g_i(\textbf{q},t) := \frac{q_i exp(- \lambda_i t)}{\sum_{j=1}^{n} q_j exp(- \lambda_j t) }, i = 1,..., n
\label{eq10}
\end{equation}

Thus, the conditional residual off-time is also hyper-exponentially distributed with parameters \textbf{$\lambda$} and $g(\textbf{q},t)$, with $g(\textbf{q},t)$ being the n-tuple of functions $g_i(\textbf{q},t)$ for $i \in [1,n]$.
 
\section{MDP-based model} 
\label{Section4}
Markov Decision Processes (MDP), also referred to as stochastic dynamic programs or stochastic control problems, are a mathematical framework for sequential decision making when the outcomes are uncertain \cite{Ref4}. \\

The elements of the MDP are: 
\begin{itemize}
\item \textbf{Decision epochs: }they correspond to $T_k$, the time denoting the beginning of each sleep period $B_k$ until achieving a terminal state representing the end of a generic idle period $I$. We are considering then discrete decision epochs and a finite horizon MDP. 
\item \textbf{State space: }at each decision point, the system space can be represented by $\textbf{q}$, the current probability distribution of the residual off-time. The initial state is denoted $\textbf{q}^0$, the probability distribution of the total off-time. At each stage, the probability distribution of the residual off-time is updated through the operator $\textbf{g}$ given in Equation (\ref{eq10}). The probability distribution at time $t_k$ is $\textbf{q} = \textbf{g}(\textbf{q}^0,t_k)$. The state space is then the set of the different probability vectors \textbf{q} which can be infinite. We limit ourselves to a finite set $Q$. 
\item \textbf{Action space: }it corresponds to the different possible sleep levels. 
\item \textbf{Transition probabilities: } the probability of going from state \textbf{q} to state \textbf{q'} after going to sleep level $l$ is given by $P_{\textbf{q},l,\textbf{q'}} = \mathbbm{1}_{\textbf{q'}=g(\textbf{q},l)}$ where the operator $\mathbbm{1}$ takes the value 1 if the condition is verified and 0 otherwise. \\
\end{itemize}

\begin{figure} [ht]
\centering
\includegraphics[width=8.8cm,height= 5.6cm]{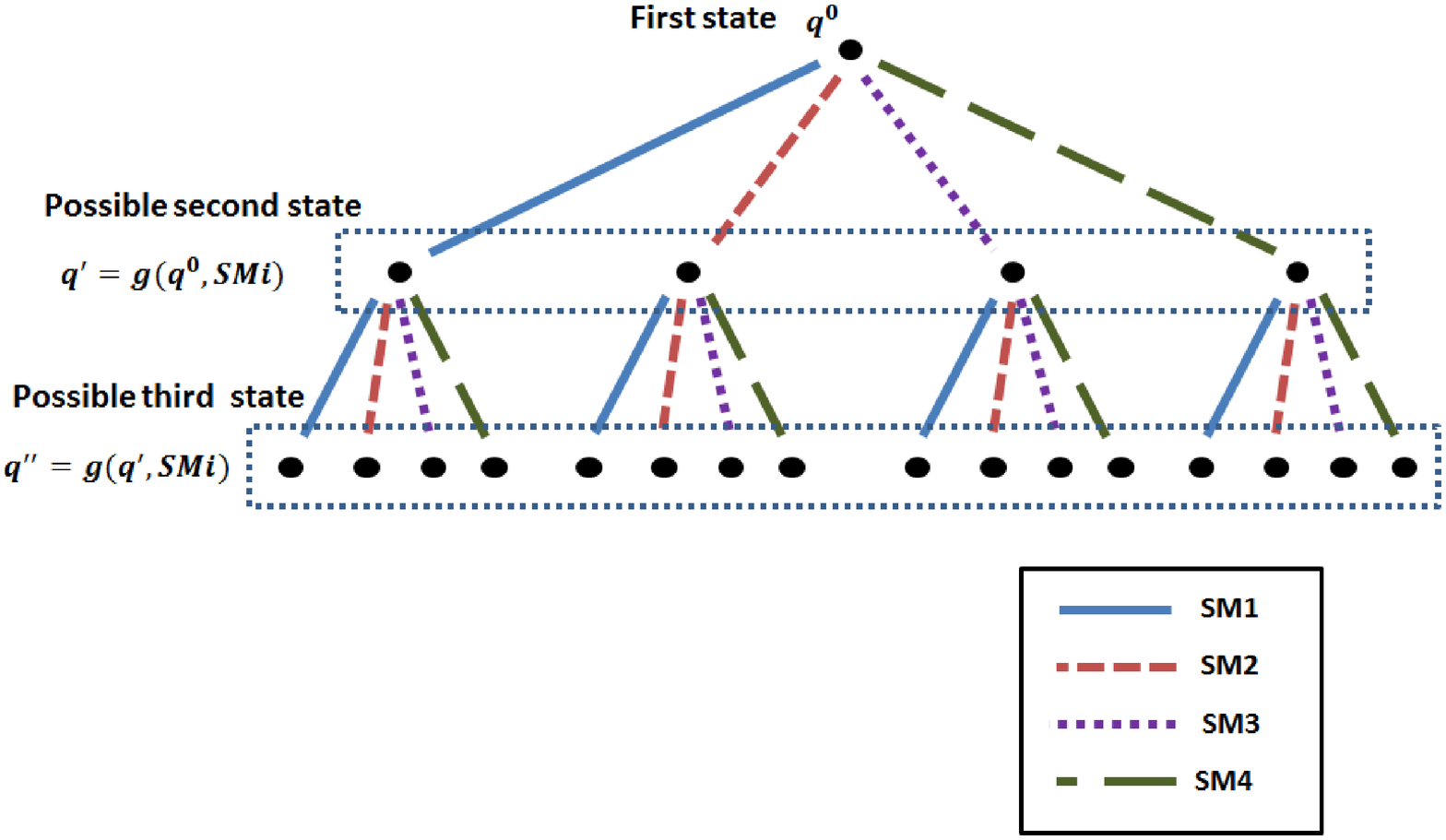}
\caption {Possible states and transitions during an idle period} 
\label {Tree}
 \end{figure}

Figure \ref{Tree} illustrates this formulation. We start in the initial state \textbf{q}$^0$ (first circle). We choose an action corresponding to SM level $l$ among the four ASM choices. This leads us to the second state $\textbf{q'} = \textbf{g}(\textbf{q}^0,t_{SM})$ where $t_{SM}$ is the duration of the chosen SM. Starting from \textbf{q'}, we again have four possible choices and so on. In order to have a finite state space, one can limit the depth of the tree to a certain level. For a given $\tau$, this depth can be computed as the number of times we can repeat the lowest sleep mode level until we cover all the inactivity period of length $\tau$. 
 
\subsection{Dynamic Programming solution}
Dynamic Programming (DP) is a collection of algorithms enabling to compute an optimal policy when the model is perfectly known, as is the case in MDP \cite{Ref5}. \\

Following \cite{Ref1}, we introduce the following dynamic programming: 
\begin{equation}
V_k^*(t_k) = \min_{b_{k+1} \ge 0} \{\mathbb{E}[c(t_k,b_{k+1})] + P(\tau_{t_k} > b_{k+1})V_{k+1}^*(t_{k+1})\}
\end{equation}

where: 
\begin{itemize}
\item $ V_k^*(t_k)$ is the optimal cost at time $t_k$.
\item $t_k$ corresponds to the observation point at the end of a sleep period  $B_{k}$.
\item $b_{k+1}$ denotes the duration of the next sleep period $B_{k+1}$.
\item $\tau_{t_k} $ is the conditional residual off-time at time $t_k$.
\item $c(t_k,b_{k+1})$ is the stage cost at $t_k$ when the control is $b_{k+1}$.
\item $P(\tau_{t_k} > b_{k+1})$ represents the transition probability. \\
\end{itemize}  

The per stage cost can be written as follows: 
\begin{equation}
c(t,b) = \epsilon_1 \mathbb{E}[(b-\tau_t)\mathbbm{1}_{\tau_{t} \leq b}] + \epsilon_2 P_l b
\end{equation}  
where $\mathbb{E}[(b-\tau_t)1_{\tau_{t} < b}]$ and $P_l b$ are both normalized values. 

As in \cite{Ref1}, we can replace the subscript $t_k$ in the cost function $V$ by the current state $\textbf{q}$. 
Note that $V$ is defined for a given initial state. DP allows us to solve $V$ for any initial state. \\

The DP algorithm can then be written as: 
\begin{equation}
V(\textbf{q}) = \min_{b \ge 0} \{\mathbb{E}[c(\textbf{q},b)] + P(\tau(\textbf{q}) > b)V(g(\textbf{q},b))\}
\end{equation}
with the cost function being changed as: 
\begin{equation}
c(\textbf{q},b) = \epsilon_1 \mathbb{E}[(b-\tau(\textbf{q}))\mathbbm{1}_{\tau(\textbf{q}) \leq b}] + \epsilon_2 P_l b 
\end{equation}  

The delay can be computed as in \cite{Ref1}, as follows: 
\begin{equation}
\mathbb{E}[(b-\tau(\textbf{q}))\mathbbm{1}_{\tau(\textbf{q}) \leq b}] = b - \sum_{i=1}^{n} q_i \frac{1 - exp(- \lambda_i b) }{\lambda_i }
\end{equation}
\subsection{Value iteration algorithm}

In order to solve the DP, we make use of the following Value iteration algorithm.

\begin{table}[h]
\begin{tabular}{p{8.5cm}}
  \hline
  \hline  \vspace{0.1cm}
  \textbf{Value Iteration Algorithm}  \vspace{0.1cm}  \\
  \hline
\hline \\
Initialize $V(\textbf{q}) = 0, \forall  \textbf{q} \in Q$ \\~\\

Repeat
\begin{equation}
        \begin{aligned}
        & \Delta \leftarrow 0 \\
        & \text{For each } \textbf{q} \in Q \\
        & \hspace{20pt} v \leftarrow V(\textbf{q}) \\
        & \hspace{20pt} V_{k+1}(\textbf{q}) = \min_{b \ge 0} \{c(\textbf{q},b) + P(\tau(\textbf{q}) > b)V_k(g(\textbf{q},b))\} \\
        & \hspace{20pt} \Delta \leftarrow \max(\Delta, |v-V(\textbf{q})|) \\
        \end{aligned}   
\end{equation}

until $\Delta < \eta$ (small positive number)\\~\\

Output a deterministic policy $\pi$, such that: \\~\\

\hspace{33pt} $\pi(\textbf{q}) = \argmin_b \{c(\textbf{q},b) + P(\tau(\textbf{q}) > b)V(g(\textbf{q},b))\}$  \\~\\
\hline
\hline
\end{tabular}
\end{table}

Proofs of convergence are given in \cite{Ref1}.

\subsection{Accounting for switching cost}
With the formulation developed so far, the optimal policy can yield oscillations between the different sleep levels within the same idle period. These oscillations can be costly in practice as switching to a sleep mode and waking up from it costs energy too. In order to be able to account for this extra energy term and be able to reduce oscillations, we add another variable $\beta$ to the cost function denoting the switching cost. $\beta$ can be written as follows: 
\begin{equation}
\label{eq1}
\beta = \begin{cases} 1 & \mbox{if next sleep level is different from current one} \\~\\
0 &\mbox{otherwise }
\end{cases}
\end{equation}

Thus, the per stage cost becomes: 
\begin{equation}
\label{eq17}
c(\textbf{q},b) = \epsilon_1 \mathbb{E}[(b-\tau(\textbf{q}))\mathbbm{1}_{\tau(\textbf{q}) \leq b}] + \epsilon_2 P_l b + \epsilon_3 \beta
\end{equation}  
where $\epsilon_3$ is the weight given to the switching cost. We have $\sum_{i=1}^3 \epsilon_i = 1$.  

\section{Numerical applications}
\label{Section5}
\subsection{System configuration}
We consider a base station with the following configuration: 2x2 MIMO, Radiated Power: 46 dBm, Bandwidth: 20 MHz. For the sake of illustration, we consider in this section only SM levels 2 and 3, their powers consumptions are described in Table \ref{Table2}. They are computed using IMEC power model tool \cite{IMEC} for the present configuration. The idle state is when the base station is activated but not transmitting anything.

 \begin{table}[ht]
    \renewcommand{\arraystretch}{1.3}
\caption{Power consumption values in Watts}
    \label{Table2}
\centering
    \begin{tabular}{|c|c|c|c|}
    \hline
	\hspace{0.1cm} \textbf{Active} \hspace{0.1cm} & \hspace{0.1cm} \textbf{Idle} \hspace{0.1cm} & \hspace{0.1cm} \textbf{SM$_2$} \hspace{0.1cm} & \hspace{0.1cm} \textbf{SM$_3$} \hspace{0.1cm}  \\ 
    \hline 
    250 & 109 & 14.3 & 9.51 \\
    \hline
    \end{tabular}
    \end{table}  
    
 The parameters of the hyper-exponential distribution are taken as $\lambda = [10, 500]$ and $q = [\frac{1}{2}, \frac{1}{2}]$.
  
 \subsection{Optimal ASM policy without switching cost}

\begin{figure}[ht]
    \centering
    \begin{subfigure}[b]{0.475\textwidth}
        \centering
        \includegraphics[width=\textwidth]{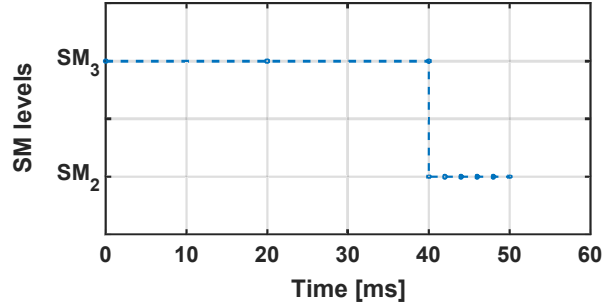}
        \caption{$\epsilon_1 = 0.3$}
        \vspace{10pt}

    \end{subfigure}
    \begin{subfigure}[b]{0.475\textwidth}  
        \centering 
        \includegraphics[width=\textwidth]{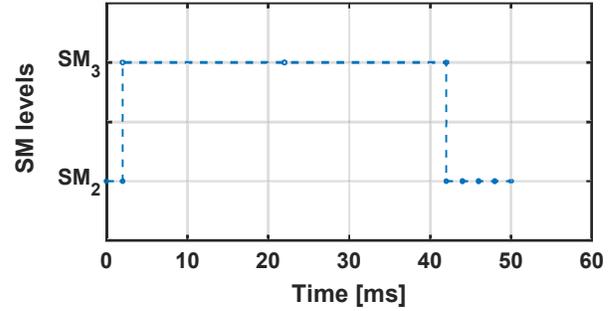}
        \caption{$\epsilon_1 = 0.7$}
    \end{subfigure}
    \vspace{10pt}

    \begin{subfigure}[b]{0.475\textwidth}   
        \centering 
        \includegraphics[width=\textwidth]{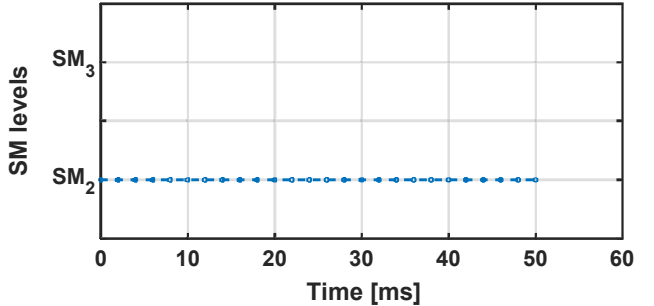}
        \caption{$\epsilon_1 = 1$}
    \end{subfigure}
\caption {Policies achieved by the Value iteration algorithm after convergence} 
\label {Fig3}
\end{figure}

Figure \ref{Fig3} presents the policy achieved at the limit of the Value iteration algorithm (after convergence) when $\epsilon_1$ takes three different values: \{0.3, 0.7, 1\}. The x-axis representing the time corresponds to one generic inactivity period. For $\epsilon_1 = 0.3$, the system values energy reduction more than delay reduction so the optimal policy is to choose the deepest sleep mode, SM$_3$, whenever is possible. For $\epsilon_1 = 0.7$, the delay reduction is more valued than the energy reduction, hence the system chooses first SM$_2$ then because of the heavy tailed nature of the off-time period, it does not expect an arrival to take place soon, it hence switches to SM$_3$ and then back to SM$_2$ so that it is in the lightest SM when an arrival is about to happen. For $\epsilon_1$ equal to 1, the delay is prioritized hence the system chooses only SM$_2$. 
 
\begin{figure} [ht]
\centering
\includegraphics[width= 8.8cm,height= 4.5cm]{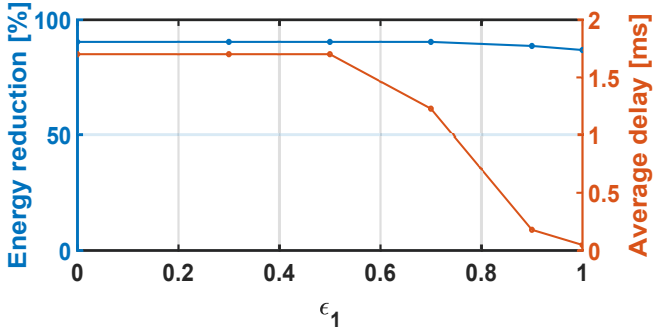}
\caption {Energy reduction and delay induced by the different policies} 
\label {Fig4}
 \end{figure}

Figure \ref{Fig4} shows both the energy reduction achieved by the chosen policy for each $\epsilon_1$ and the delay that it incurred. When $\epsilon_1$ is very low, we reach high energy savings (up to 90.4\%) with an average delay around 1.7$ms$. The more we increase $\epsilon_1$, the more weight we give to the delay. Thus, fewer levels of sleep will be used which induces a decrease in the energy savings but also less impact on the delay. When $\epsilon_1 = 1$, only the lowest SM level (SM$_2$) is chosen. In this case, the average delays are negligible and the energy reduction reaches 86.88\%. Note that the power consumption of the illustrative SM$_2$ and SM$_3$ levels are not very different, as shown in Table \ref{Table2}, which explains that the decrease in the energy reduction shown in Figure \ref{Fig4} is not very large. 
%

\subsection{Adding switching cost}
Figure \ref{Fig3} shows that some policies, for instance when $\epsilon_1 = 0.7$, may present oscillations between different sleep levels which justifies in this case the use of the switching cost $\beta$ in order to get more stable policies. 

We show in this section the impact of the switching cost $\beta$ for a given configuration. As an example, we take $\epsilon_1$ = 0.7, we vary $\epsilon_3$ and fix $\epsilon_2 = 1 - (\epsilon_1+\epsilon_3)$. \\

\begin{figure}[ht]
    \centering
    \begin{subfigure}[b]{0.475\textwidth}
        \centering
        \includegraphics[width=\textwidth]{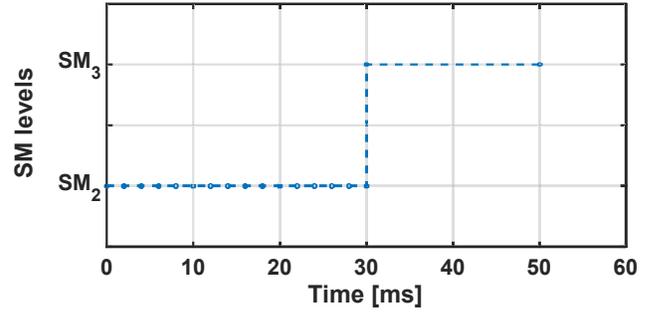}
        \caption{$\epsilon_3 = 0.1$}
    \end{subfigure}
    \vskip\baselineskip
    \begin{subfigure}[b]{0.475\textwidth}  
        \centering 
        \includegraphics[width=\textwidth]{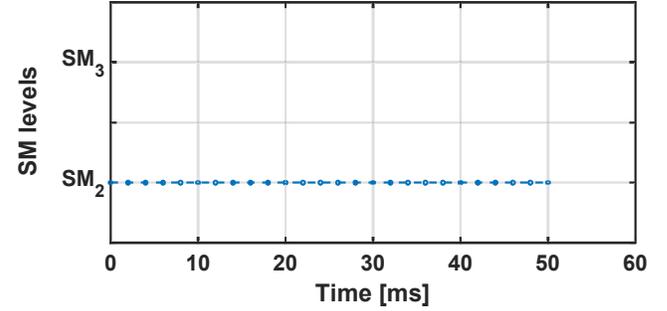}
        \caption{$\epsilon_3 = 0.2$}
    \end{subfigure}
\caption {Different Policies depending on $\epsilon_3$ when $\epsilon_1 = 0.7$} 
\label {Fig6}
\end{figure}

Figure \ref{Fig6} shows the variation of the policy depending on the values given to $\epsilon_3$. We can see that the more we increase it, the more grouped the sleep levels become. 
 
\begin{figure} [ht]
\centering
\includegraphics[width= 8.8cm,height= 4.5cm]{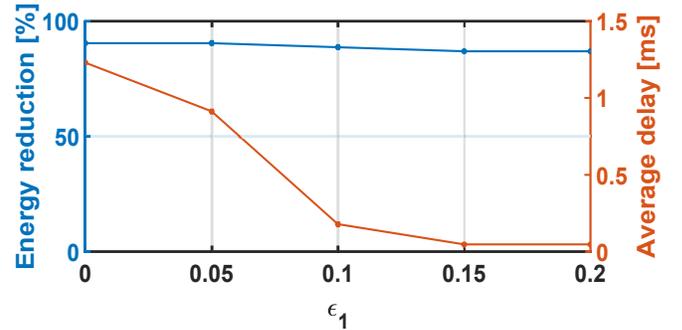}
\caption {Variation of the energy reduction and the average delay depending on the weight of the switching cost} 
\label {Fig7}
 \end{figure}

Figures \ref{Fig7} shows the energy reduction and the delay as a function of the weight $\epsilon_3$ given to the switching cost. The more $\epsilon_3$ is increased, the less energy savings and average delays are attained. This can be explained by the fact that the system begins by the lowest sleep levels in the first states. When we impose the switching cost, the system tends not to switch from these first levels which in turn reduces the potential energy savings. 

\section{Conclusion}
\label{Section6}
We presented in this work optimal control strategies enabling to implement the ASMs in 5G NSA. Based on an MDP approach, different policies can be derived according to the trade-off between energy consumption reduction versus the delay incurred by the sleep levels. The corresponding cost function has to take into consideration the policy of the network operator regarding the thresholds imposed on the delay and/or the energy savings. In a latency-sensitive use case such as Ultra reliable Low Latency Communications (URLLC) for instance, the weight put on the delay must be the highest possible. In more delay-tolerant scenarios, a tradeoff is needed. We also studied the impact of adding a switching cost in order to be able to reduce the oscillations between the sleep levels during a single idle period. This strategy stabilises the control and forces the system to remain in the first sleep levels. Here too, a trade-off between system stability versus performance is to be set by the network operator.

As an extension to this work, we aim to study a system based on 5G SA architecture and take the signaling periodicity into account in the decision process. This periodicity can be either fixed or can come as an outcome of the control strategy.

\end{document}